\documentclass[traditabstract]{aa}
\usepackage[varg]{txfonts}
\usepackage{graphicx}
\usepackage{xcolor}
\usepackage{natbib}
\usepackage{amsmath,amssymb}
\usepackage{hyperref}
\usepackage{rotating, bm}
\usepackage{ulem}

\newcommand{\mgii}{\ion{Mg}{ii}}
\newcommand{\BLOS}{$B_{\rm LOS}$}

\begin{document} 

\title{Magnetic field diagnostics of prominences with the \ion{Mg}{ii}\,k line\\ 3D Stokes inversions versus traditional methods}

\titlerunning{Prominence diagnostics with Mg\,{\sc ii}\,k}

\author{
Ji\v{r}\'{\i} \v{S}t\v{e}p\'an\inst{1}
\and
Tanaus\'u del Pino Alem\'an\inst{2,3}
\and
Javier Trujillo Bueno\inst{2,3,4}
}

\authorrunning{\v{S}t\v{e}p\'an et al.}

\institute{
Astronomical Institute of the Czech Academy of Sciences, Fri\v{c}ova 298, 25165 Ond\v{r}ejov, Czech Republic
\and
Instituto de Astrof\'{\i}sica de Canarias, E-38205 La Laguna, Tenerife, Spain
\and
Departamento de Astrof\'{\i}sica, Universidad de La Laguna, E-38206 La Laguna, Tenerife, Spain
\and
Consejo Superior de Investigaciones Cient\'{\i}ficas, Spain
}

\date{Received XXXX; accepted XXXX}
 
\abstract{
The Mg\,{\sc ii}\,k resonance line is commonly used for diagnosing the solar chromosphere. We theoretically investigated its intensity and polarization in solar prominences, taking 3D radiative transfer and Hanle and Zeeman effects into account. We used an optically thick 3D model representative of a solar prominence and applied several inversion methods to the synthetic Stokes profiles, clarifying their pros and cons for inferring prominence magnetic fields. We conclude that the self-consistent 3D inversion with radiative transfer is necessary to determine the magnetic field vector, although its geometry cannot be inferred with full fidelity. We also demonstrate that more traditional methods, such as those based on the weak field approximation or the constant-property slab assumption, can offer useful information under certain conditions.
}


\maketitle

\section{Introduction}

Among the various spectral lines used in the diagnostic of the solar chromosphere and prominences, we have the \mgii\,h and k lines, located at 280.3\,nm and 279.6\,nm, respectively. This ultraviolet (UV) resonance doublet is known for its strong sensitivity to the physical conditions of the plasma, making it a valuable tool for probing.

Recently, there has been increased interest in these UV lines thanks to the spectroscopic observations obtained by the Interface Region Imaging Spectrograph \citep[IRIS;][]{2014SoPh..289.2733D} and the spectropolarimetric data collected by the two Chromospheric LAyer SpectroPolarimeter missions, CLASP2 \citep{2016SPIE.9905E..08N} and CLASP2.1 \citep{2021AGUFMSH52A..06M}. These suborbital space experiments have provided unprecedented observations of \mgii\,h and k lines, confirming previous theoretical predictions and providing new insights into the structure and dynamics of the upper solar atmosphere \citep[see the review by][]{2022ARA&A..60..415T}.

Building on the success of these missions, the scientific community is now planning the Chromospheric Magnetism Explorer (CMEx) space telescope \citep{2023AAS...24221402B}, which will focus on the spectropolarimetry of the spectral region of the \mgii\,h\,\&\,k doublet. This new mission is expected to provide detailed information on the magnetic field and dynamics of the plasma in the chromosphere and prominences.

For the last 40 years, many studies have attempted to understand the formation of the Mg\,{\sc ii} h \& k lines in prominences to explain the available prominence observations. The first 2D freestanding slab models of prominences were developed by \citet{1982ApJ...254..780V} using the complete frequency redistribution (CRD) approximation for calculating the \mgii\,k line intensity. Later, \citet{1993A&A...274..571P} found similar results for the core of the h and k lines with both the CRD approximation and accounting for partial frequency redistribution (PRD) effects in prominences. \citet{2014A&A...564A.132H} conducted a thorough analysis using 1D prominence slab models and confirmed that the results obtained using a two-level \mgii\ atomic model without a continuum agree with those found using a multilevel plus continuum model. Their investigation indicated that prominences are generally optically thick in the k line, with line-center thicknesses reaching up to $10^3$ or $10^4$. \citet{2018A&A...618A..88J} compared 1D models with IRIS data of prominences and find that the integrated line intensities require optically thick plasma.

The magnetic field strength in quiescent solar prominences presumably ranges from a few to several tens of gauss, although it can occasionally reach significantly higher values \citep{2018RvMPP...2....1H}. While in the chromosphere the h\,\&\,k doublet exhibits remarkable linear polarization signals in the profile wings due to $J$-state interference and magneto-optical effects \citep{2012ApJ...750L..11B,2016ApJ...831L..15A,2016ApJ...830L..24D,2020ApJ...891...91D}, only the line core is visible in prominences. Therefore, our study focused on the core of the \mgii\,k line. This line is sensitive to both the Zeeman and Hanle effects. Its circular polarization degree due to the Zeeman effect scales linearly with $\mathcal{R}=1.5\times 10^{-4}B$ \citep[see Eq.~1 of][]{2022ARA&A..60..415T}, allowing for the determination of the line-of-sight (LOS) component of the magnetic field, \BLOS, while the transversal component in the plane of the sky (which scales with $\mathcal{R}^2$) would be very difficult (or impossible) to determine via the Zeeman effect at the expected magnetic field strengths. The linear polarization of the \mgii\,k line arises dominantly from the scattering of anisotropic radiation and the Hanle effect. The critical Hanle field of the line is $B_{\rm H}\approx 22\;{\rm G}$ \citep[see Eq.~2 of][]{2022ARA&A..60..415T} and is therefore suitable for quiescent prominence diagnostics. However, diagnosing magnetic fields becomes challenging when the field is not constant along the LOS and when radiation transfer needs to be considered because the prominence plasma is optically thick ($\tau>1$) and not in local thermodynamic equilibrium (LTE), that is, when it is in the so-called non-LTE regime.

The main aim of this paper is to theoretically investigate the Stokes inversion problem of the \mgii\,k line in solar prominences, taking the effects of 3D radiative transfer (RT) into account. To this end, we used an optically thick 3D prominence model with a magnetic field that varies spatially along the model's loop-like structure. Because the model's physical properties vary along the three spatial directions, the axial symmetry of the incident radiation field can break at each point within the medium without the need of a magnetic field. Such nonmagnetic causes of symmetry breaking can have an important impact on the linear polarization signals caused by the scattering of anisotropic radiation \citep[e.g.,][]{2013PORTA,2021ApJ...909..183J},  which at the line center are sensitive to the presence of magnetic fields via the Hanle effect. In addition to anisotropic radiation pumping and the Hanle effect, our study also includes the Zeeman effect, which dominates the line's circular polarization. Given that in prominences the \mgii\,k line does not show extended wings, where the effects of PRD and $J$-state interference are very important \citep{2012ApJ...750L..11B}, in this investigation we solved the 3D\,non-LTE RT problem assuming CRD without $J$-state interference. Our approach to the 3D Stokes spectral synthesis and inversion problem without assuming LTE can be found in \cite{2022A&A...659A.137S}.

Here we focused on investigating the suitability of the \mgii\,k line for spectropolarimetric diagnostics of optically thick prominences, which requires understanding the magnetic field's impact on the polarization of the emergent spectral line radiation. In particular, we ask if our 3D Stokes inversion, which consistently accounts for RT without assuming LTE, can uncover the global magnetic field geometry of an optically thick prominence using the \mgii\,k line. Section~\ref{sec:model} describes the prominence model and the spectral synthesis of the emergent Stokes profiles, taking the effects of RT  into account in the 3D model. In Sect.~\ref{sec:inv} we apply the weak field approximation \citep[WFA; see Sect.~9.6 in][]{LL04}, a Bayesian approach based on the constant-property slab model, and the full 3D Stokes inversion to the synthetic data, assessing the goodness of the inference. Finally, Sect.~\ref{sec:concl}  outlines our conclusions.

\section{Prominence model and synthesis of the Stokes profiles 
\label{sec:model}}

\begin{figure}
\centering
\includegraphics[width=0.9\hsize]{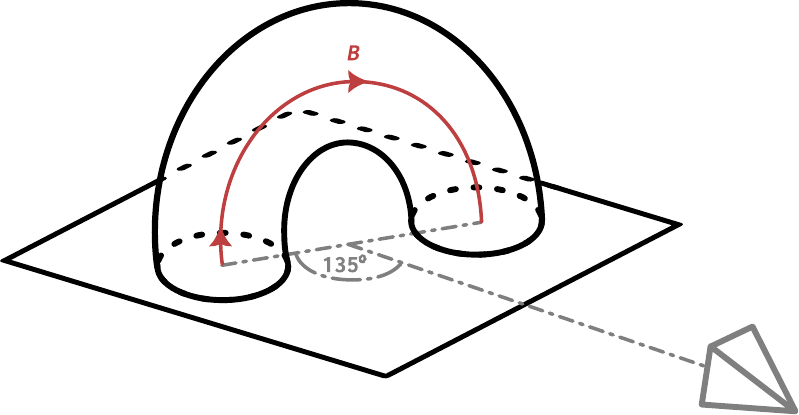}\\
\includegraphics[width=\hsize]{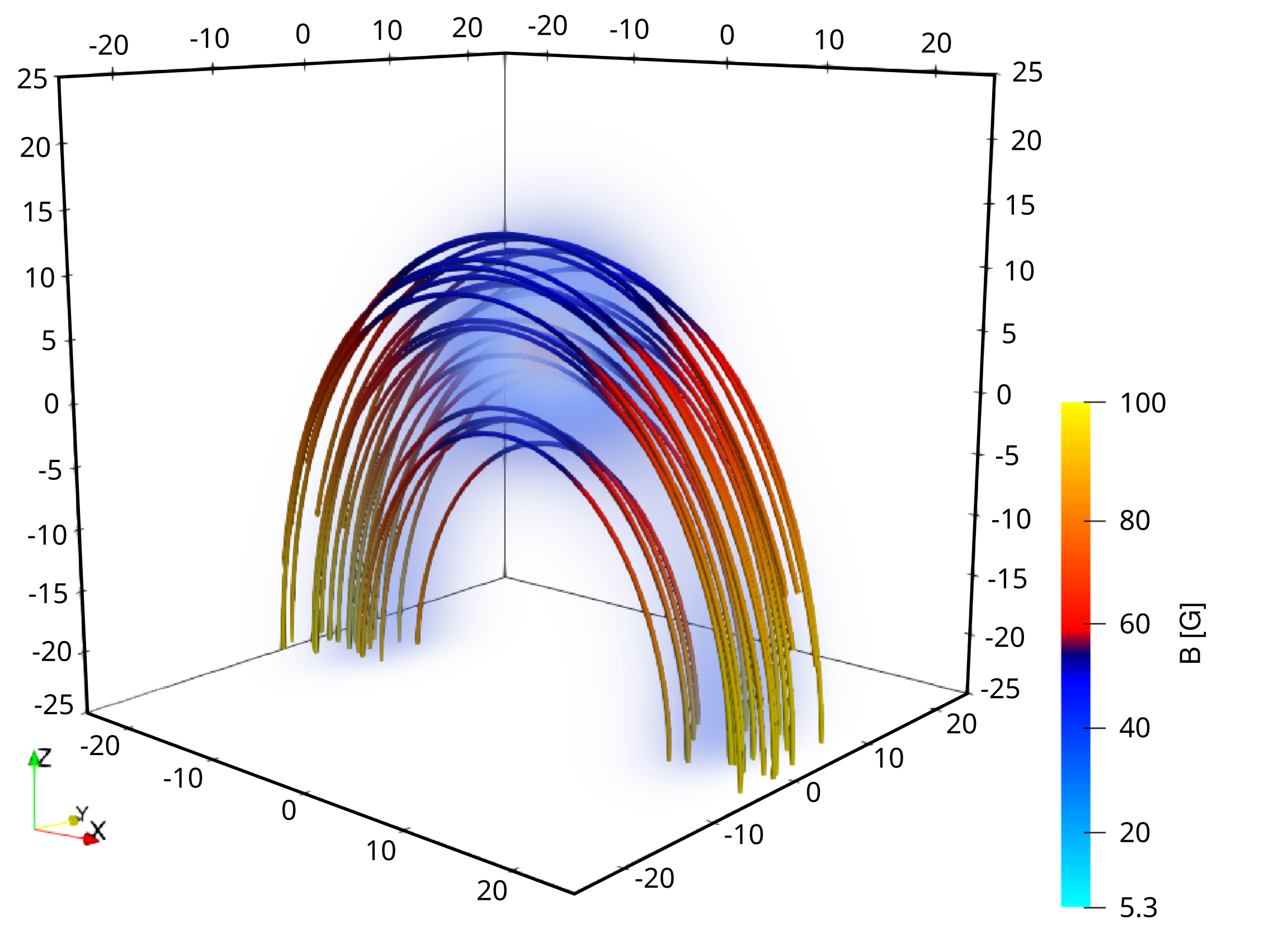}
\caption{
Geometry of the model prominence.
Top panel:
LOS of the observation. The symmetry plane of the prominence model is rotated by 45$^\circ$ with respect to the LOS. The red curve indicates the geometry of the magnetic field.
Bottom panel:
Visualization of the magnetic field lines in 
the $50\times50\times50$\,Mm$^3$ spatial domain of the 3D prominence model.
}
\label{fig:model}
\end{figure}

\begin{figure}
\centering
\includegraphics[width=\hsize]{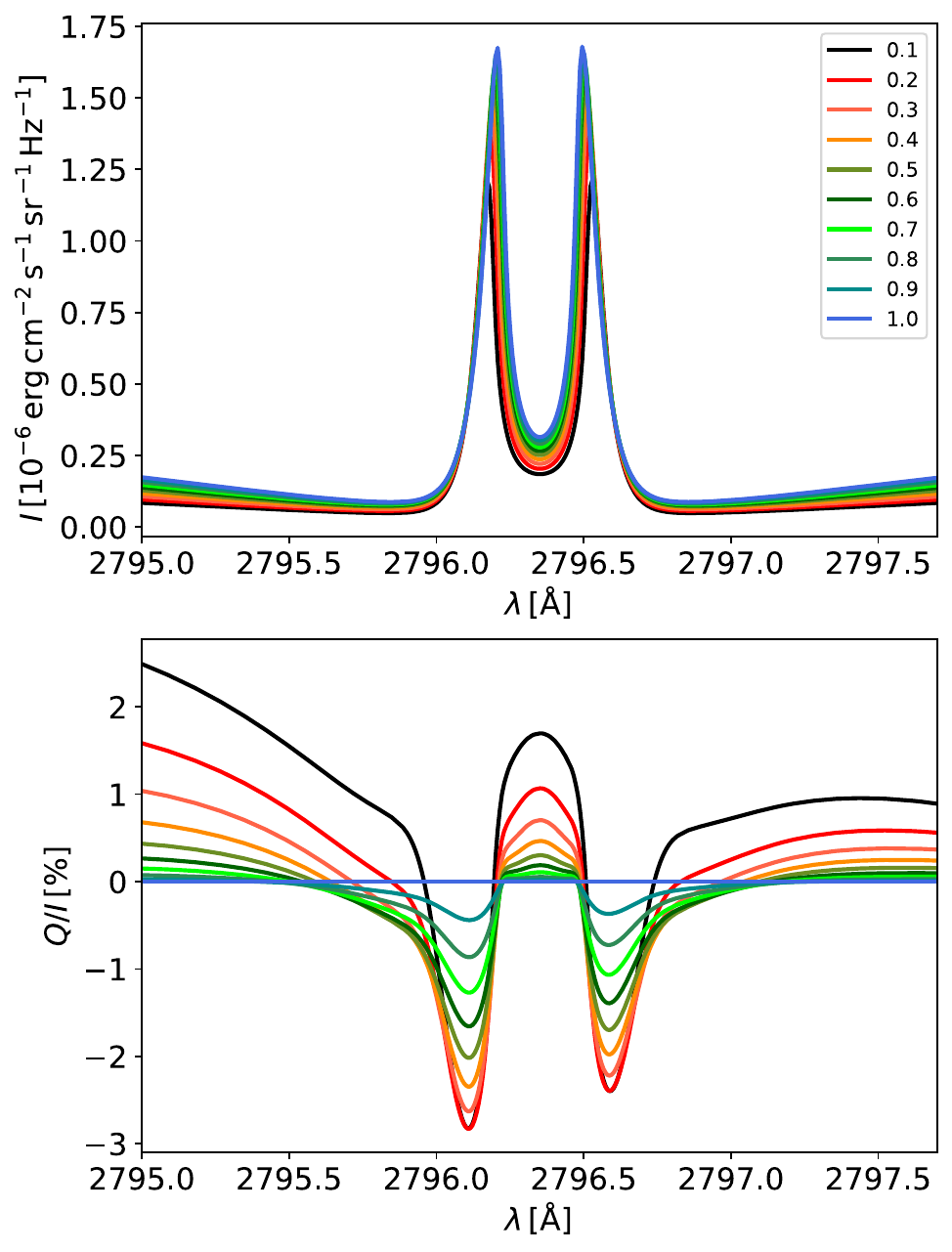}
\caption{
 Stokes $I$ and $Q/I$ profiles of the emergent \mgii\,k line radiation calculated in the FAL-C model atmosphere for various cosines ($\mu$) of the heliocentric angle, from 0.1 to 1. The Stokes $I$ and $Q$ profiles were used to obtain the illumination of the boundaries of the 3D prominence model.
}
\label{fig:falc}
\end{figure}

\begin{figure*}
\centering
\includegraphics[width=0.75\hsize]{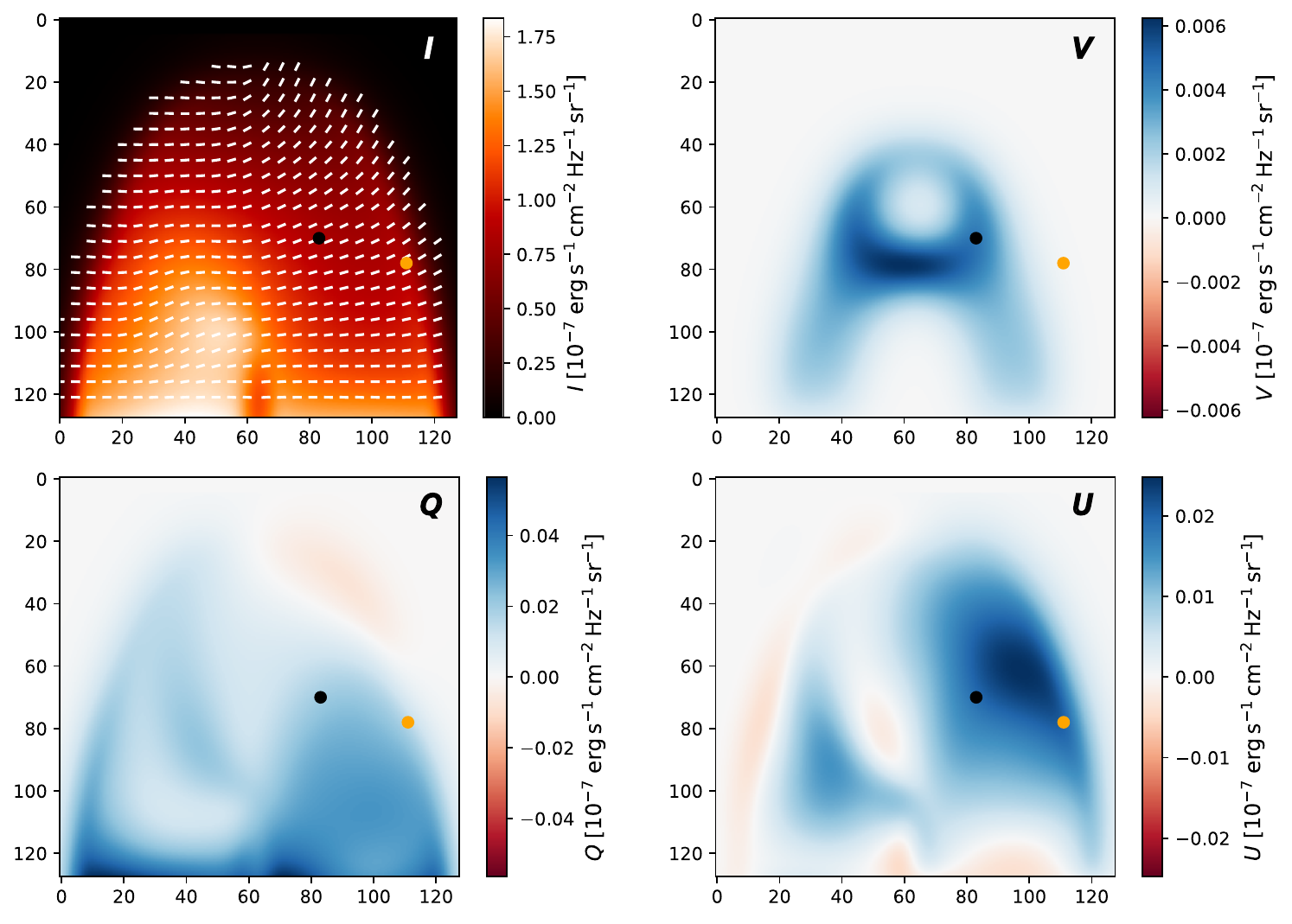}
\caption{
Emergent radiation at each point in the $128\times 128$ pixels FOV in the observation as illustrated in the top panel of Fig.~\ref{fig:model}. The $I$, $Q$, and $U$ signals are shown at the line-center wavelength, while the $V$ signal is shown at $\Delta\lambda=-0.079$\,\AA\ from the line center. The white vectors in the intensity panel show the orientation of the linear polarization at the line center. The black and orange dots indicate two particular locations in the FOV analyzed in the text.
}
\label{fig:fov}
\end{figure*}

\begin{figure*}
\centering
\includegraphics[width=0.75\hsize]{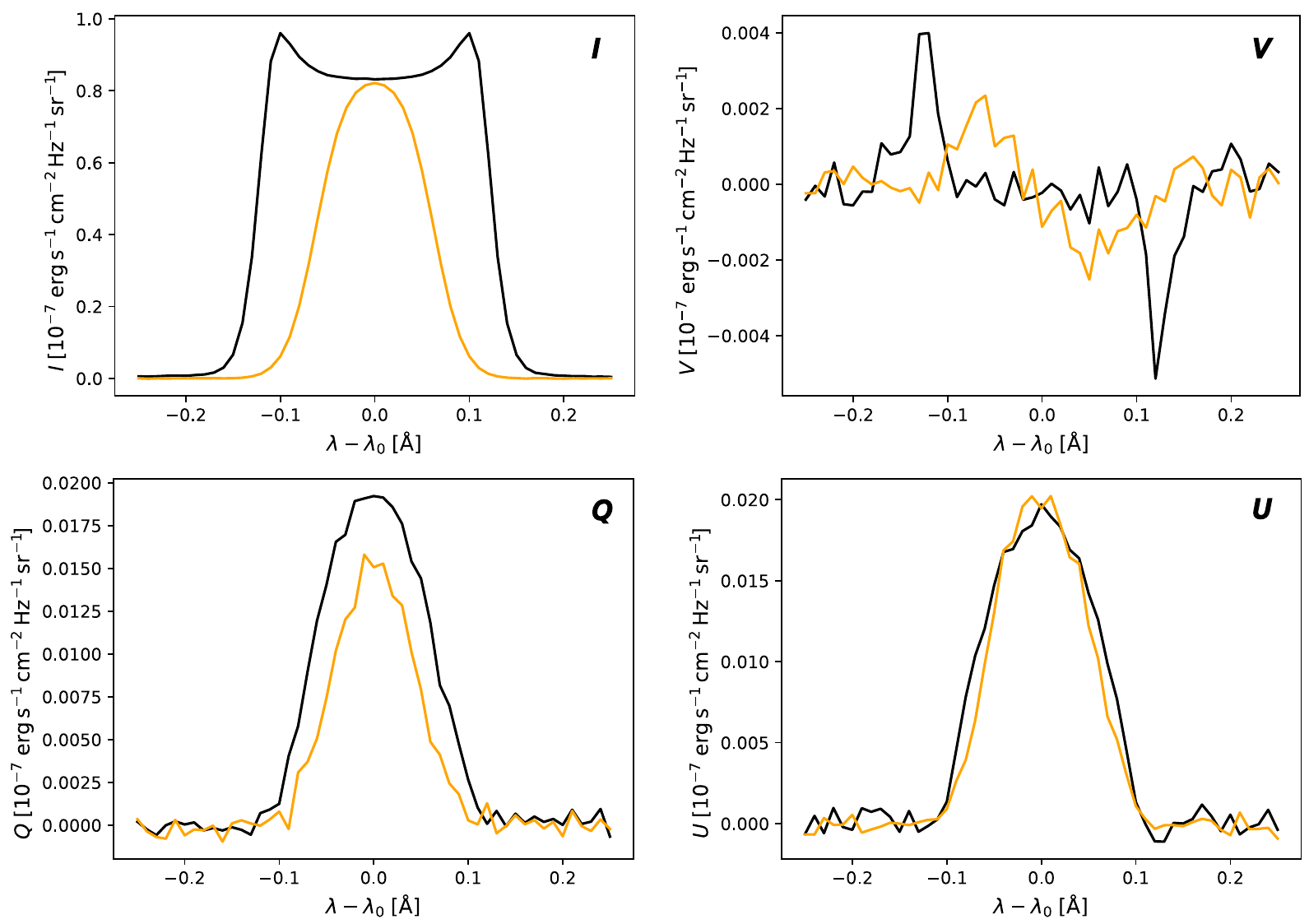}
\caption{
Stokes profiles with an added Gaussian noise with $\sigma= 5\times 10^{-11}\,{\rm erg\,s^{-1}\,cm^{-2}\,Hz^{-1}\,sr^{-1}}$ at the points indicated by the black and orange dots in Fig.~\ref{fig:fov}.
}
\label{fig:sp}
\end{figure*}

\begin{figure}
\centering
\includegraphics[width=\hsize]{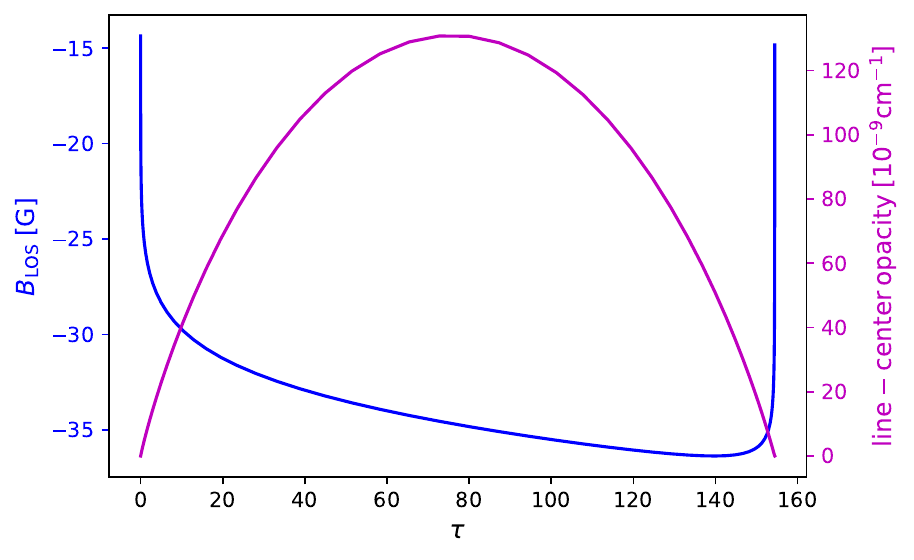}\\
\includegraphics[width=\hsize]{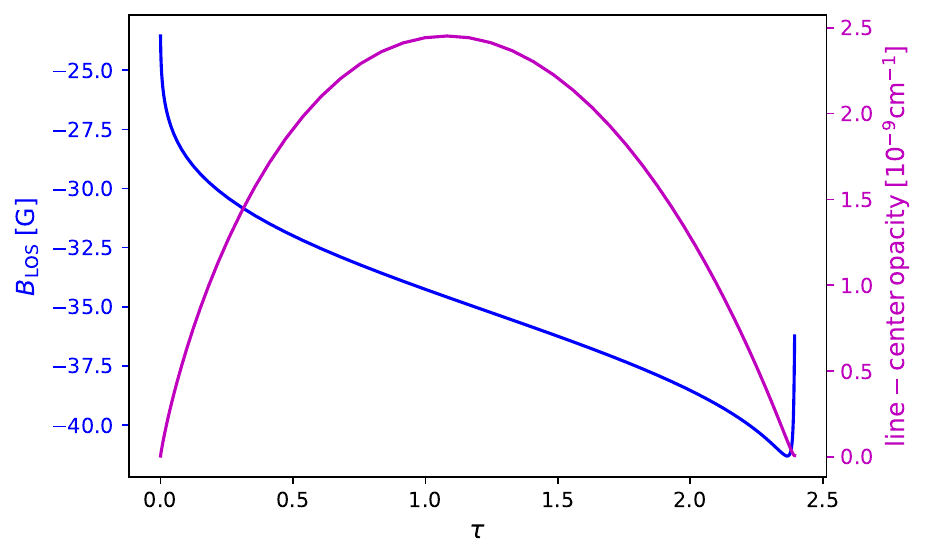}
\caption{
Variation of the LOS component of the magnetic field (blue line) and of the line-center opacity (magenta line) at the black (top panel) and orange (bottom panel) spatial points indicated in Fig.~\ref{fig:fov}. The horizontal axis gives the line-center optical depth along the LOS.
}
\label{fig:los}
\end{figure}

\begin{figure}
\centering
\includegraphics[width=0.8\hsize]{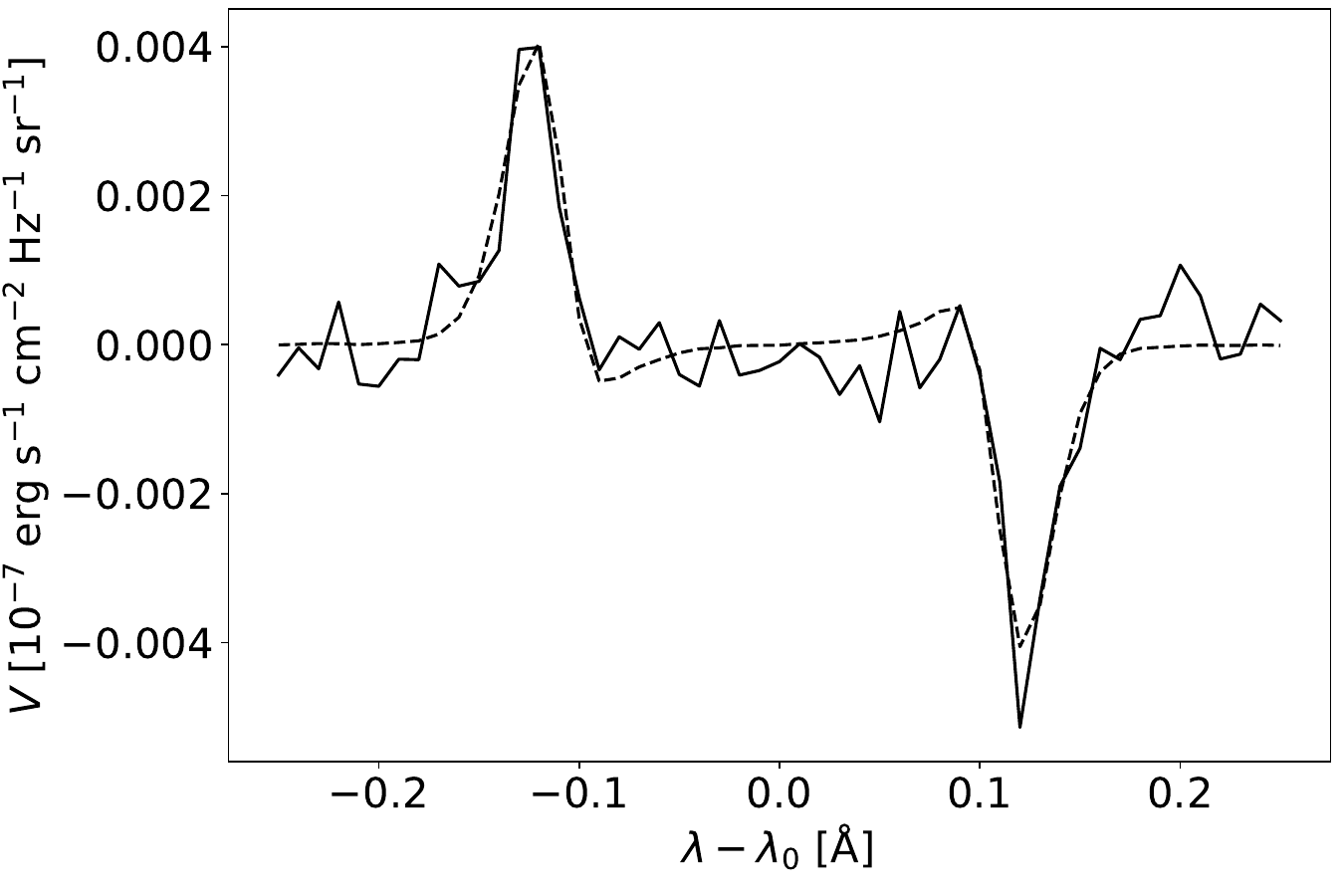}\\
\includegraphics[width=0.8\hsize]{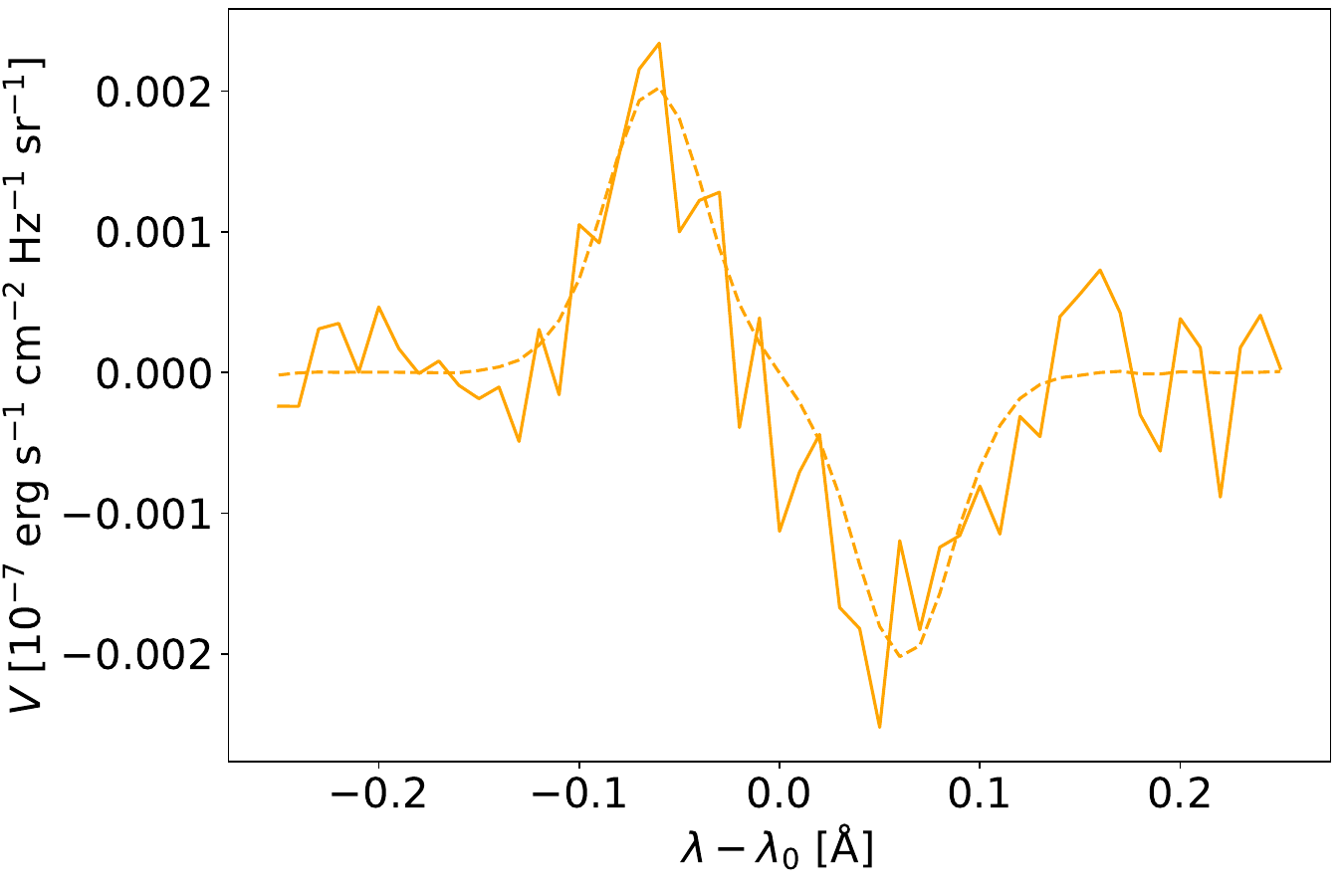}\\
\caption{
WFA best fits (dashed curves) of the spectra shown in Fig.~\ref{fig:sp} (solid curves). A noise level of $\sigma=5\times 10^{-11}\,{\rm erg\,s^{-1}\,cm^{-2}\,Hz^{-1}\,sr^{-1}}$, corresponding to approximately to $6\times 10^{-4} I_{\rm LC}$, has been added to the profiles, where $I_{\rm LC}$ is the line-center intensity. The inferred longitudinal components of the magnetic field are $-34.9\pm2.3$~G for the top panel and $-36.6\pm3.6$~G for the bottom panel.
}
\label{fig:wfa}
\end{figure}

\begin{figure*}
\centering
\includegraphics[width=\hsize]{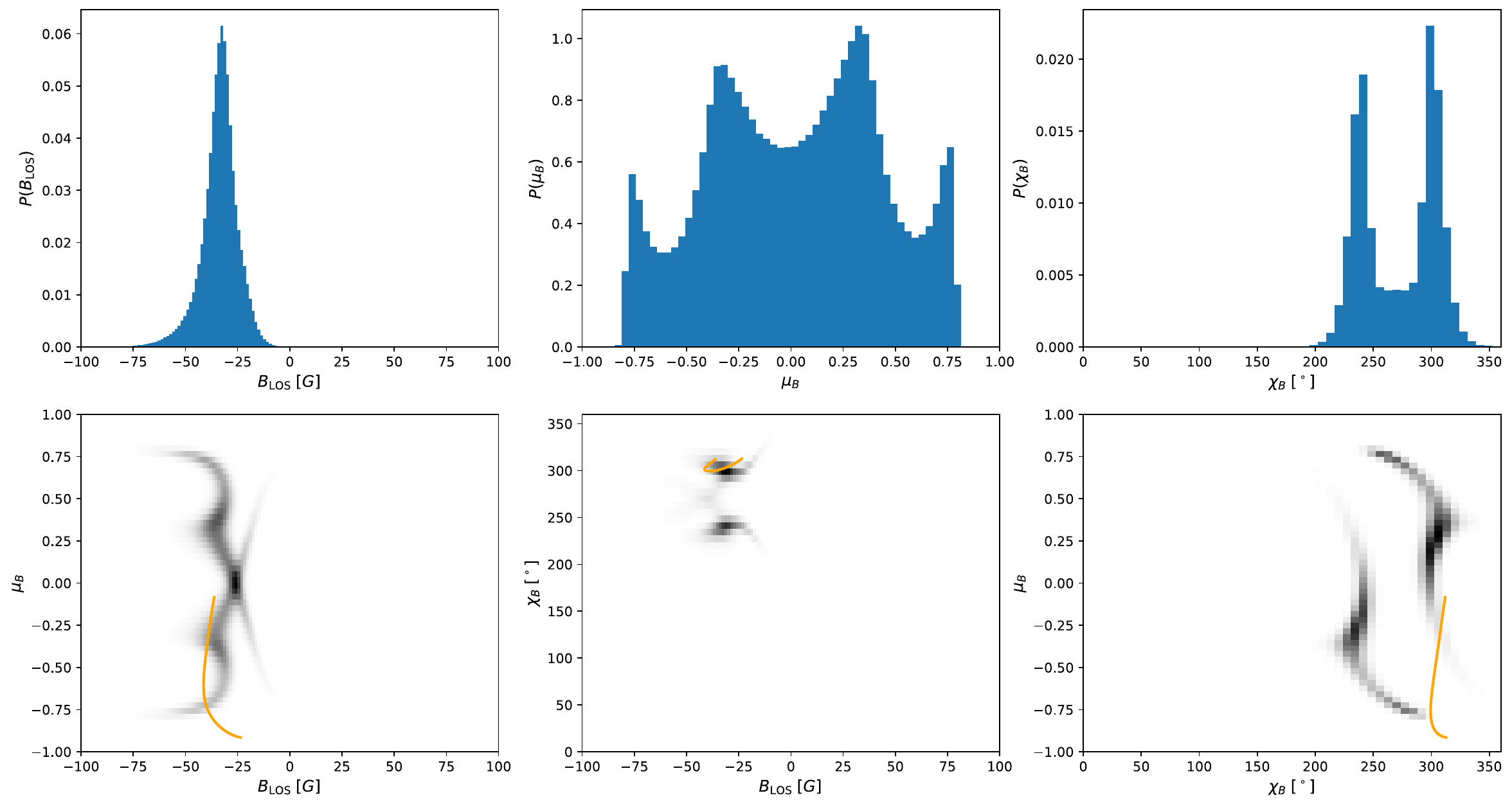}
\caption{
Marginal posteriors of the Bayesian model parameters. The orange curves in the bottom panels indicate the variation of the actual parameters along the chosen LOS in the spatial domain of the original model. See the main text for details.
}
\label{fig:post}
\end{figure*}

\begin{figure}
\centering
\includegraphics[width=0.8\hsize]{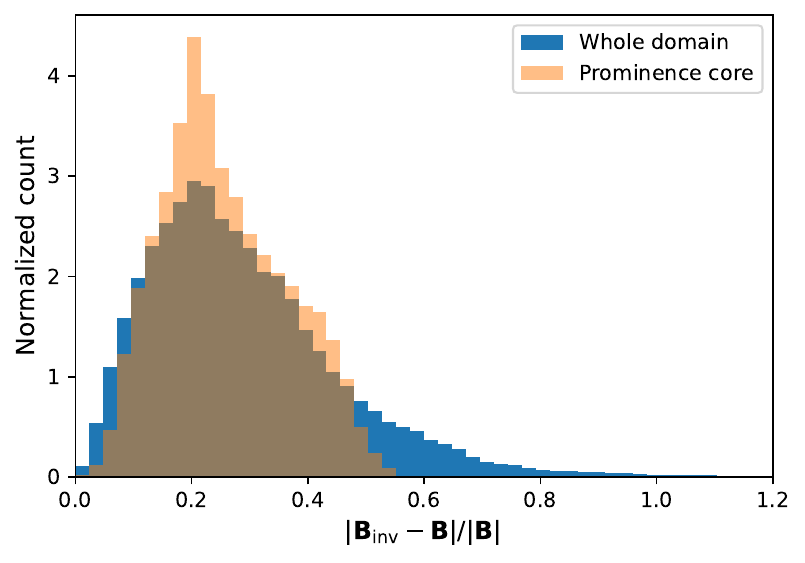}
\caption{
Histogram of the errors (see Eq.~\ref{eq:3derr})
in the magnetic field inferred via the 3D Stokes inversion.
}
\label{fig:3derr}
\end{figure}

\begin{figure*}
\centering
\includegraphics[width=0.85\hsize]{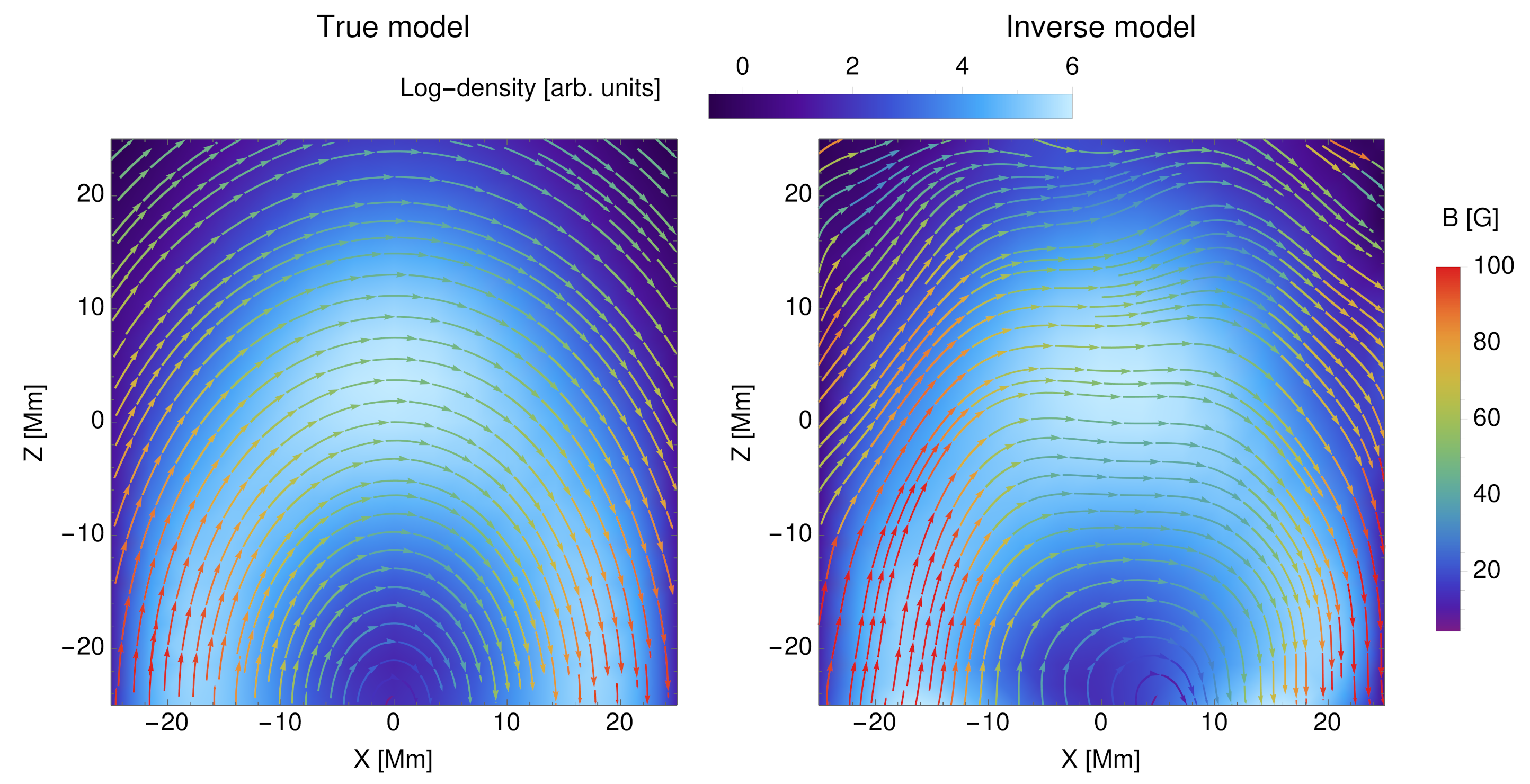}
\caption{
Comparison of vertical slices through the $Y=0$\,Mm plane in the original model and in the model resulting from the 3D inversion. The left panel shows the magnetic field lines and the density in the original model, and the right panel shows the same quantities in the inferred model. The background color in both panels represents the log-density of \mgii\ ions. Our 3D Stokes inversion successfully captures the essential morphology of the original model.
}
\label{fig:mginv}
\end{figure*}

Figure~\ref{fig:model} illustrates the prominence model we used to investigate the performance of the three inference methods we applied to the synthetic Stokes profiles. Although it is an academic prominence model with a relatively simple magnetic field geometry, it features the main ingredients that can affect the performance of the inversion, namely a nontrivial variation of the physical quantities both across the field of view (FOV) and along the LOS, and a relatively large optical thickness to manifest the effects of RT in 3D geometry.

For simplicity, we assumed that all Mg atoms are in the Mg\,{\sc ii} ionization stage. At chromospheric temperatures the Mg\,{\sc ii} ion is indeed the dominant species \citep{2013ApJ...772...89L}, and at about 15--20~kK the Mg\,{\sc ii}/Mg\,{\sc iii} fraction quickly decreases and Mg\,{\sc iii} becomes the dominant species \citep{2014A&A...564A.132H}. We modeled the Mg\,{\sc ii} k resonance line using a two-level atom model and in the limit of CRD. These two approximations can be justified because we can directly specify the total population in the lower and upper levels of the k line of Mg\,{\sc ii} (avoiding the need to account for an equation of state and for ionization and recombination processes) and because the line center, dubbed k$_3$, is not strongly affected by PRD effects \citep{2013ApJ...772...89L,2017SSRv..210..183T}.\footnote{While strong velocity gradients can make the line center sensitive to PRD effects due to the Doppler shifts \citep{2017A&A...597A..46S}, the lack of bright wings in prominence profiles may diminish this effect.} Finally, we assumed an abundance of $7.544$ for magnesium. Given the academic nature of our 3D prominence model, we assumed that hydrogen, which is the most important donor of electrons under the considered thermodynamic conditions, is fully ionized. We could then estimate the electron volume density directly from the volume density of magnesium and its abundance. The inelastic collisional rates in the \mgii\,k line transition have been calculated using the data from \citet{1995JPhB...28.4879S}.

The spatial domain of our prominence model is $50\times50\times50$~Mm$^3$; it contains a loop-like structure whose legs start at the top of the chromosphere and extends up to about 40~Mm above it; the spatial domain sits on top of the C model of \citealt{1993ApJ...406..319F}, hereafter the FAL-C model, at about 2.2~Mm above the visible solar surface. The Mg\,{\sc ii} number density is prescribed and decreases outward from the inner part of the prominence body. The optical thickness is maximum at the central part of the apex of the loop-like structure and exceeds 100 at the \mgii\,k line center. The prominence is isothermal, with a temperature of 10~kK, with a microturbulent velocity of 5~km/s \citep{2014A&A...564A.132H}. For simplicity, we do not include bulk velocities in this model. Finally, the magnetic field is such that the field lines follow the loop-like structure (see Fig.~\ref{fig:model}), with a strength reaching about 100~G close to the chromosphere and decreasing to about 40~G at the apex. These relatively strong magnetic fields have been chosen to produce a circular polarization signal above the considered noise level.

The prominence is illuminated by the underlying chromosphere, not limited to the computational domain. This illumination is axially symmetric and only depends on the angle between the propagation direction and the local vertical. The spectral profiles and their center-to-limb variation (see Fig.~\ref{fig:falc}) have been calculated with HanleRT-TIC\footnote{The 1D non-LTE code is publicly available at \url{https://gitlab.com/TdPA/hanlert-tic}.} \citep{2016ApJ...830L..24D} using the FAL-C model. We note that in our solution we did not approximate the geometry of the solar chromosphere by an infinite plane, but we took the curvature of the solar surface into account.

We solved the non-LTE\,RT problem in 3D by applying the spectral synthesis mode of our code \citep{2022A&A...659A.137S}, obtaining the $J^K_Q$ radiation field tensor components everywhere in the domain, and the emergent Stokes profiles for the chosen LOS (see Fig.~\ref{fig:model}). The resulting FOV is shown in Fig.~\ref{fig:fov}. The linear polarization is due to the scattering of anisotropic radiation and the Hanle effect, while the circular polarization is caused by the Zeeman effect. For the application of the inference methods presented in Sect.~\ref{sec:inv}, we selected two particular positions in the FOV (black and orange dots in Fig.~\ref{fig:fov}). The spectra corresponding to these two locations, after adding Gaussian polarimetric noise with $\sigma=5\times 10^{-11}$~erg\,s$^{-1}$ cm$^{-2}$\,Hz$^{-1}$\,sr$^{-1}$, on the order of $5\times10^{-4}$ of the line-center intensity, are shown in Fig.~\ref{fig:sp}. The material behind the black dot LOS has an optical thickness at the line center of about 160, showing the typical self-reversal in its intensity profile, while the orange dot LOS has a more modest optical thickness of about 2.5 at the line center, showing no self-reversal. In Fig.~\ref{fig:los} we show how \BLOS\ and the line center opacity change with the optical thickness at the line center, $\tau$, along the LOS for these two selected positions.

\section{Magnetic field inference
\label{sec:inv}}

In this section we apply three methods to infer the magnetic field in the prominence model described in Sect.~\ref{sec:model}. In Sects.~\ref{ssec:wfa} and \ref{ssec:bayes} we apply two commonly used inference techniques, both of them based on strong assumptions for the RT. These two methods are applied pixel by pixel, and for the presentation of the results we chose two particular positions in the FOV (see the black and orange dots in Fig.~\ref{fig:fov}), which are representative of a LOS with significant optical thickness at the line center ($>100$) and a LOS with a modest optical thickness at the line center ($\sim2.5$). In Sect.~\ref{ssec:3dinv} we apply our 3D Stokes inversion to infer the magnetic field vector in the whole prominence body and its surroundings. In particular, we are interested in assessing the goodness of the fits in the different methods and discussing their advantages and disadvantages.

\subsection{Weak field approximation (WFA)
\label{ssec:wfa}}

When the Zeeman splitting of the line's levels produced by the magnetic field is much smaller than the spectral line width, and under other certain assumptions, it is possible to find a closed and simple expression for the circular polarization, the so-called WFA \citep[see, e.g., Sect.~9.6 of][]{LL04},
\begin{equation}
V(\lambda) = -C\lambda^2_0B_{\rm LOS}g_{\rm eff}\frac{\partial I}{\partial\lambda}(\lambda),
\label{eq:wfa}
\end{equation}
where $C = 4.6686\times 10^{-13}$~G$^{-1}$\AA$^{-1}$, $\lambda_0$ is the wavelength of the spectral line in \AA, $g_{\rm eff}$ is the effective Land\'e factor ($7/6$ for the \mgii\,k line), and \BLOS\ is the longitudinal component of the magnetic field in gauss.

The first assumption that must be fulfilled for the applicability of the WFA is that the Zeeman effect must be the only mechanism contributing to the polarization. While in solar prominences the linear polarization of the Mg\,{\sc ii} k line is produced by the scattering of anisotropic radiation and the Hanle effect, the circular polarization is dominated by the Zeeman effect. Consequently, in this subsection we focus only on the circular polarization profiles.

The second assumption for the applicability of the WFA is that \BLOS\ must be constant along the LOS. While this condition seems really restrictive and close to impossible to fulfill, in practice it means that \BLOS\ must be approximately constant along the LOS in those regions actually contributing to the emergent profiles. It is also clear from Eq.~\eqref{eq:wfa}, which relates Stokes $V$ and the derivative of the intensity $I$, that when the profiles are formed in extensive regions, the contributions to both Stokes parameters must come from the same regions along the LOS.

If we assume uncorrelated and Gaussian noise, we can use Eq.~\eqref{eq:wfa} to calculate \BLOS\ as follows \citep{2012MNRAS.419..153M},
\begin{equation}
B_{\rm LOS} = -\frac{1}{C}
\frac{\sum_j V(\lambda_j)I'(\lambda_j)}{\sum_j(I'(\lambda_j))^2} \pm 
\frac{\sigma}{C\sqrt{\sum_j(I'(\lambda_j))^2}},
\label{eq:wfa-blos}
\end{equation}
where $I'(\lambda_j) = \lambda^2_0g_{\rm eff}\frac{\partial I}{\partial\lambda}(\lambda_j)$,
$\lambda_j$ are the observed wavelengths, $\sigma$ is the standard deviation of the Gaussian distribution of the noise in Stokes $V$, and the error is computed from the covariance matrix assuming a wavelength-independent standard deviation and a confidence level of 68.3\,\% (1$\sigma$).

In Fig.~\ref{fig:wfa} we show the WFA fit to the Mg\,{\sc ii} k circular polarization profiles shown in Fig.~\ref{fig:sp}, corresponding to the black and orange dots in Fig.~\ref{fig:fov}, by applying Eq.~\eqref{eq:wfa-blos}. The fits are quite good and the corresponding \BLOS\ are $-$34.9$\pm$2.3 and $-$36.6$\pm$3.6~G. From Fig.~\ref{fig:los}, we see that the retrieved \BLOS\ correspond approximately to the magnetic field around the maximum of the line-center opacity. We note that the $\tau$ scale in Fig.~\ref{fig:los} corresponds to the line center wavelength and that the opacity quickly decreases for the line wings. Therefore, even if at the line core we cannot ``see'' the regions with $\tau\sim 80$ in Fig.~\ref{fig:los}, at the near wing wavelengths we can.

The WFA has the advantage of providing an estimation of \BLOS\ with negligible computing effort given that in our 3D model the required conditions are satisfied. However, the WFA returns a single number and an uncertainty that only accounts for the noise; there is thus no information about the magnetic field geometry and stratification, nor about other sources of uncertainty in the inferred values. For the particular cases we studied in this work, it turns out that \BLOS\ is relatively constant along the formation region of the profiles, keeping the same polarity along the whole LOS, and thus the inferred \BLOS\ are rather good estimates. However, for more complex magnetic field geometries, with potential cancellation effects (different polarities along the LOS) or strong source function gradients, the WFA does not guarantee a good estimation of \BLOS.

\subsection{Constant-property slab Bayesian inversion
\label{ssec:bayes}}

Most of the magnetic field inference in prominences (and filaments) over the last two decades relies on the modeling of spectropolarimetric observations in the He\,{\sc i} triplet at 10830~\AA\ and D$_3$ triplet at 5877~\AA\ assuming a constant-property slab illuminated by the solar radiation of the underlying quiet Sun disk \citep[see, e.g., the review by][and references therein]{2022ARA&A..60..415T}. Several inference methods based on this model can be found in the literature, such as look-up tables based on principal component analysis \citep[e.g.,][]{2003ApJ...598L..67C}, minimization methods such as that implemented in the Hanle and Zeeman light (HAZEL) code \citep{2008ApJ...683..542A}, or through Bayesian statistical approaches \citep[e.g.,][]{2019A&A...625A.128D}.

One of the main assumptions of this modeling approach is that the radiation pumping within the slab is fully dominated by the cylindrical symmetric illumination from the underlying solar disk (i.e., that the excitation of the atoms within the slab is not affected by RT within the slab). In order for this approximation to be reasonable, the optical thickness of the slab plasma should be small 
enough \citep{2007ApJ...655..642T,2023A&A...675A..45V}. The model then assumes that all properties of the plasma within the slab are constant along the LOS. Typically, a single slab is assumed in the modeling, but several components -- both side by side \citep[e.g.,][]{2010A&A...520A..77X} and one after the other along the LOS \citep[e.g.,][]{2012ApJ...759...16M} -- have been considered.

When the modeling assumptions are satisfied, these inference methods can provide estimations of the magnetic field vector while allowing for the study of ambiguities and uncertainties. However, when the optical depth is on the order of, or larger than, the unity along any direction within the prominence (or filament), or if the plasma properties are not constant in the region along the LOS where the line forms, the accuracy of the inference can be severely compromised. Moreover, the inference methods based on Bayesian statistics, while providing a clear picture of the uncertainties and ambiguities, are computationally heavy, especially when considering more than a single slab.

In prominences, the Mg\,{\sc ii} k line investigated in this work typically shows larger optical thickness than the He\,{\sc i} triplet lines  \citep{2018A&A...618A..88J}. Consequently, an unsuitable performance of the constant-property slab model is, a priori, expected. At the pixel marked with a black dot in Fig.~\ref{fig:fov} the plasma of our 3D model is very optically thick at the k-line center, with an optical depth of over one hundred. Its intensity profile (black curve in Fig.~\ref{fig:sp}) shows a clear self-reversal, which cannot be reproduced assuming a single constant-property slab. On the contrary, at the orange dot pixel in Fig.~\ref{fig:fov} the plasma of our 3D model has a total optical depth of about $2.5$ along the LOS. This pixel is near the prominence ``edge'' and it can thus ``see'' most of the underlying chromosphere. We note, however, that the prominence body blocks some of the chromospheric radiation, so the assumption of cylindrically symmetric illumination is not fully valid. The \BLOS\ is approximately constant along the LOS for the orange dot pixel (see the bottom panel of Fig.~\ref{fig:los}), but the inclination of the magnetic field vector changes along the LOS.

We calculated the Bayesian posterior distribution for the case of a single-component constant-property slab model inversion of the emergent Stokes profiles at the location of the orange dot pixel. The parameters of our model are (i) \BLOS, (ii) the cosine of the polar angle of the magnetic field inclination with respect to the solar radius ($\mu_B$), (iii) its azimuth in the plane normal to the radius with respect to the projection of the LOS on such a plane ($\chi_B$), (iv) the thermal width of the line ($\Delta v_D$), and (v) and the line-center optical thickness ($\tau$). We used a Jeffreys prior\footnote{We performed identical calculations using a uniform prior for $\tau$ and the results we have obtained are very similar. Nevertheless, our numerical experiment shows that the uniform prior for $\tau$ leads to slight overestimation of both $\tau$ and \BLOS.} for $\tau$ and uniform priors for the rest of the parameters, between 0 and 1000~G for \BLOS, between -1 and 1 for $\mu_B$, between 0 and $2\pi$ for $\chi_B$, and between 0.1 and 10~km/s for $\Delta v_D$.

Figure~\ref{fig:post} shows the marginalized posterior distributions for \BLOS, $\mu_B$, and $\chi_B$. Although the physical properties of the 3D model at the selected pixel do not fulfill the applicability conditions, because the optical depth is larger than unity, the illumination is not cylindrically symmetric, and $\mu_B$ changes along the LOS, the inferred \BLOS\ turns out to be as good as with the WFA (see Sect.~\ref{ssec:wfa}) and, moreover, the constant-property slab approach is capable of finding $\chi_B$ up to the ambiguities. This inversion method is much slower than the WFA, but in exchange it provides additional physical information.

Even though the \BLOS\ inference is rather good, the magnetic field strength is overestimated by about a factor of 2 (hence the magnetic field energy density by a factor of 4) due to the significant uncertainty in $\mu_B$. Due to the symmetry assumed in the 3D model, there are ambiguous solutions for both $\mu_B$ and $\chi_B$.

For an optically thin prominence with a not-too-complex magnetic field geometry, approaches based on this constant-property slab model (Bayesian inference, principal component analysis, etc.) seem to be optimal since they can provide maximum information on uncertainties even when there are no self-consistent RT constraints.

\subsection{3D Stokes inversion
\label{ssec:3dinv}}

The third inference method we applied to the synthetic Stokes profiles that we calculated by solving the non-LTE\,RT problem in our 3D prominence model is the 3D Stokes inversion approach described in \citet{2022A&A...659A.137S}. This method approaches the inverse problem by finding the physical quantities in the whole spatial domain of the 3D model simultaneously. Different regions of the model's spatial domain are coupled by the transfer of polarized radiation and, in addition, the solution can have additional constraints such as those from the magneto-hydrodynamic equations. As described in detail in the aforementioned paper, this mesh-free method does not rely on calculating a sequence of self-consistent forward models that lead to the minimum of a merit function. Instead, it follows an unconstrained minimization method in which unphysical solutions are allowed but penalized via regularization terms in the merit function. This allows relatively accurate solutions to be obtained within a significantly shorter computing time. This method does not only provide a solenoidal magnetic field vector $\vec B$ everywhere in the model's spatial domain, but for our particular case also other thermodynamic quantities such as the atomic number density.

We solved the inversion problem using 480~CPU cores of the OASA computer of the Astronomical Institute in Ond\v{r}ejov. The solution shown here was reached in about 20~hours or $10^4$~CPU hours. As mentioned above, we penalized unphysical (non-self-consistent) solutions and magnetic field vector distributions not fulfilling $\nabla\cdot\vec B=0$. The initial state of the magnetic field vector components has been chosen so that all the amplitudes of the basis functions were randomly sampled from a normal distribution with zero mean and a standard deviation of 20\,G. The initial guess of the atomic number density has been a constant function equal to $\log_{10}N=1$ in the units of $[N]=\mathrm{cm^{-3}}$.

To evaluate the goodness of the inversion, we quantified the error in the inferred magnetic field with
\begin{equation}
e = \frac{\|\vec B_{\rm inv}-\vec B\|}{\|\vec B\|}\,,
\label{eq:3derr}
\end{equation}
where $\vec B$ is the true magnetic field vector (see Fig.~\ref{fig:model}) and $\vec B_{\rm inv}$ is magnetic field vector resulting for the inversion. Figure~\ref{fig:3derr} shows the histogram of this quantity evaluated at $10^5$ randomly located points in the whole 3D domain (blue histogram) and with the same number of random points within the prominence body (orange histogram). As the Stokes parameters of the \mgii\,k line give us information about the ``visible'' surface of the prominence, whose size is comparable to that of the model's domain, the two histograms are relatively close.

We find a typical relative error in the inferred magnetic field vector of about 20--30\,\%. However, it is important to emphasize that using this relative error as a measure of the quality of the inference can be misleading. For instance, a small spatial displacement of a magnetic loop, for example at position $(X,Z)=(0,-25)\,{\rm Mm}$, can lead to a very significant relative error. In Fig.~\ref{fig:mginv} we show a cut in the $X$--$Z$ plane of the original and inferred spatial distribution of the magnetic field vector and of the atomic number density, demonstrating that the inversion does a pretty good job in recovering the overall physical model.

Given the available CPU time, we performed about two dozen inversions with different initializations of the model and different setups of the inversion algorithm. This is not enough to make any quantitative conclusions about the ability of different setups to perform the inversion. But overall, we have found that the model has (i) converged to a solution similar to the one presented in this section, (ii) ended up in a local minimum of the merit function, or (iii) has diverged. This last situation has occurred in cases where the inversion algorithm setup can be considered too aggressive (i.e., too large iteration steps, etc.). Importantly, the solution did never converge to any solution fundamentally different and ambiguous from the one presented here. This is indicative that, at least for this 3D academic prominence model, the global consistency imposes constraints strong enough to remove the ambiguities present in approaches assuming either no coupling between the FOV pixels or unrealistic symmetric conditions. Consequently, the nontrivial spatial coupling through RT seems to impose strong constraints on the solutions that are possible, leading to robust results.

Concerning the linear polarization signals caused by scattering processes, it is important to emphasize that they are sensitive mostly to regions not much deeper than optical depth unity along the LOS, where the anisotropy of the radiation can be substantial so as to produce observationally relevant linear polarization signals. Given that the Mg\,{\sc ii} k line can be very optically thick in this prominence model, the inversion cannot be expected to perfectly recover the magnetic field in these optically inaccessible regions for which the Stokes profiles do not give us enough information. Nevertheless, the 3D inversion provides a sufficiently good estimate of the global structure and strength of the magnetic field.

\section{Discussion and conclusions
\label{sec:concl}}

We solved the non-LTE problem of the generation and transfer of polarized radiation in the Mg\,{\sc ii} k line in an academic 3D prominence model, where the magnetic field lines follow an optically thick loop-like structure. We chose this relatively simple geometry to facilitate the comparison between different magnetic field inference approaches. We have found that, despite 3D RT effects, for this relatively simple prominence model both the WFA and the constant-property slab approaches can give good estimations of the longitudinal component of the magnetic field from the observed Stokes~$V$ profile. To some degree, the constant-property slab approach is able to estimate the magnetic field transversal component from the linear polarization. To recover the full 3D picture, a full Stokes inversion method that includes the effects of RT in 3D is necessary, such as the one we applied in this study.

We have found that the WFA provides a fast estimation of \BLOS\, that is suitable at least for a simple structure such as our prominence model. For profiles with small enough optical thickness along the LOS, methods based on the constant-property slab model seem to be the best approach in terms of recovered information and computational time required, if the prominence has a simple enough geometry. When the optical thickness exceeds unity, we can take advantage of the RT coupling by applying our full 3D RT approach and inferring the global structure of the prominence.

Due to the significant optical thickness in our 3D model, also found in actual prominences \citep[e.g.,][]{2018A&A...618A..88J}, the Stokes $Q$ and $U$ profiles are mostly sensitive to the outermost layers of the prominence (in the direction toward the observer). This is more critical in prominences because, unlike on-disk observations, the wings are not observed. In contrast with the \ion{He}{i} lines more popularly used in prominence diagnostics, the \mgii\,k line is a strong resonance line with a spectral structure. This entails that the chromospheric radiation illuminating the prominence plasma is sensitive to velocities, introducing frequency shifts between the absorption profile and the incoming illumination spectrum, and to variations in the chromospheric surface that produce significant changes in the intensity of this line.

The significant optical thickness can be even more problematic if the real physical scenario is that of many small-scale threads with their own prominence-corona transition regions \citep[e.g.,][]{2007A&A...472..929G}. In this case, the inversion can become extremely challenging. Therefore, spectral lines with smaller optical thickness in prominences, which provide more spatially averaged information on the magnetic field, may provide very valuable additional information.

The scientific importance of developing a space telescope that enables routine spectropolarimetric observations in the near-UV region of the \mgii\,h and k lines cannot be overestimated, because the polarization signals that the combined action of scattering processes and the Hanle and Zeeman effects introduce in this spectral region encode a wealth of information on the magnetism and geometry of chromospheric and prominence plasmas. Equally important is the development of advanced plasma diagnostic techniques capable of providing reliable information on the magnetic field vector.

\begin{acknowledgements}
J.\v{S}. acknowledges the financial support from project \mbox{RVO:67985815} of the Astronomical Institute of the Czech Academy of Sciences.
T.P.A.'s participation in the publication is part of Project RYC2021-034006-I, funded by MICIN/AEI/10.13039/501100011033, and the European Union
``NextGenerationEU''RTRP. T.P.A. and J.T.B. acknowledge support from the Agencia Estatal de Investigación del Ministerio de Ciencia, Innovación y Universidades (MCIU/AEI) under grant ``Polarimetric Inference of Magnetic Fields'' and the European
Regional Development Fund (ERDF) with reference
PID2022-136563NB-I00/10.13039/501100011033.
\end{acknowledgements}

\bibliographystyle{aa}
\bibliography{ms.bib}

\end{document}